\documentclass[a4paper]{article}

\usepackage{INTERSPEECH2022}
\usepackage{kotex, comment}
\usepackage{cite, url}
\usepackage{tabularx}
\usepackage{hyperref}
\usepackage{makecell}
\usepackage{multirow}
\usepackage[dvipsnames]{xcolor}
\hypersetup{
    colorlinks=true,
    linkcolor=Cerulean,
    filecolor=purple,      
    urlcolor=Cerulean,
}

\title{Two Methods for Spoofing-Aware Speaker Verification: Multi-Layer Perceptron Score Fusion Model and Integrated Embedding Projector}
\name{Jungwoo Heo, Ju-ho Kim, Hyun-seo Shin}
\address{School of Computer Science, University of Seoul, Republic of Korea}
\email{jungwoo4021@gmail.com, wngh1187@naver.com, gustjtls123@naver.com}

\begin{document}

\maketitle

\begin{abstract}
\label{section:abstract}
The use of deep neural networks (DNN) has dramatically elevated the performance of automatic speaker verification (ASV) over the last decade. 
However, ASV systems can be easily neutralized by spoofing attacks. 
Therefore, the Spoofing-Aware Speaker Verification (SASV) challenge is designed and held to promote development of systems that can perform ASV considering spoofing attacks by integrating ASV and spoofing countermeasure (CM) systems. 
In this paper, we propose two back-end systems: multi-layer perceptron score fusion model (MSFM) and integrated embedding projector (IEP). 
The MSFM, score fusion back-end system, derived SASV score utilizing ASV and CM scores and embeddings. 
On the other hand, IEP combines ASV and CM embeddings into SASV embedding and calculates final SASV score based on the cosine similarity. 
We effectively integrated ASV and CM systems through proposed MSFM and IEP and achieved the SASV equal error rates 0.56\%, 1.32\% on the official evaluation trials of the SASV 2022 challenge. 
\end{abstract}


\section{Introduction}
\label{section:introduction}
Automatic speaker verification (ASV) aims to verify whether the speaker of input utterance is identical to the enrollment utterance's speaker. 
Since deep neural networks (DNN) based speaker feature extractor achieved state-of-the-art performance in the research community, deep learning became a dominant approach to ASV study \cite{variani2014deep, dehak2010front, snyder2018x, thienpondt2021idlab, heo2020clova, shim2021graph}. 
Due to the nature of ASV task, however, the ASV system can be easily neutralized against spoofing attacks because the speaker feature extractor focused on extracting speaker information even for spoofed speech \cite{wang2020asvspoof}. 

The ASV spoof challenge series \cite{wu2015asvspoof, wu2017asvspoof, wang2020asvspoof, delgado2021asvspoof} has been held to promote spoofing countermeasure (CM) research for spoofing attack detection \cite{wang2020asvspoof}. 
These challenges provided a large-scale dataset that consists of bonafide, synthesized, and converted speech and evaluated proposed systems using minimum tandem decision cost function (t-DCF) \cite{kinnunen2020tandem} that reflects the effect of CM on the ASV system. 
With ASV spoof 2019, the best systems showed EERs less than 2\% in logical access, but it is unclear how this system benefits the ASV system because the challenge uses a fixed ASV system in t-DCF calculating. 
Therefore, some researchers have begun to consider constructing an integrated system that optimizes and evaluates ASV and CM simultaneously \cite{jung2022sasv}. 

Since the ASV and CM systems developed independently, research is required to incorporate them. 
The Spoofing-Aware Speaker Verification (SASV) challenge 2022 \cite{jung2022sasv}, a special session in ISCA INTERSPEECH 2022, is held first time this year to encourage the research of integrated systems of ASV and CM. 
Challenge aims to separate utterances 2 class as follow:
\begin{itemize}
    \item \textit{Target}: bonafide and vocalized from target speaker. 
    \item \textit{Non-target}: spoofed or vocalized from non-target speaker. 
\end{itemize}
The challenge is described in more detail in chapter \ref{section:sasv2022}.

In other studies, the back-end system approach has been widely adopted to process the output of pre-trained front-end systems \cite{eskimez2018front, kolboek2016speech, heymann2017beamnet, bai2021universal}. 
This idea has the advantage of reusing pre-trained systems and is applicable to SASV, since the ASV and CM task results are related to SASV task. 

To this end, we propose two back-end systems: Multi-layer perceptron Score Fusion Model (MSFM) and Integrated Embedding Projector (IEP), which depicts ASV and CM systems as front-end subsystems. 
MSFM is a score fusion back-end system that utilizes subsystem scores and embeddings and outputs SASV scores directly. 
Otherwise, IEP is an embedding fusion back-end system that aggregates subsystem embeddings into SASV embedding and calculates SASV score based on the cosine similarity. 
With the proposed fusion strategies, we reported superior performance compared to the SASV challenge baseline systems and ranked 5th in the challenge. 

We summarize our contribution to the SASV challenge 2022: 
(i) We proposed two back-end systems that can employ pre-trained ASV and CM systems without modification or training. 
(ii) We achieved equal error rates (EER) of 0.56\% and 1.32\% for the SASV evaluation protocol using proposed MSFM and IEP system. 
(iii) Finally, proposed MSFM system placed 5th in the challenge.

\begin{figure*}[ht!]
\begin{center}
    \centering
    \includegraphics[width=0.8\linewidth]{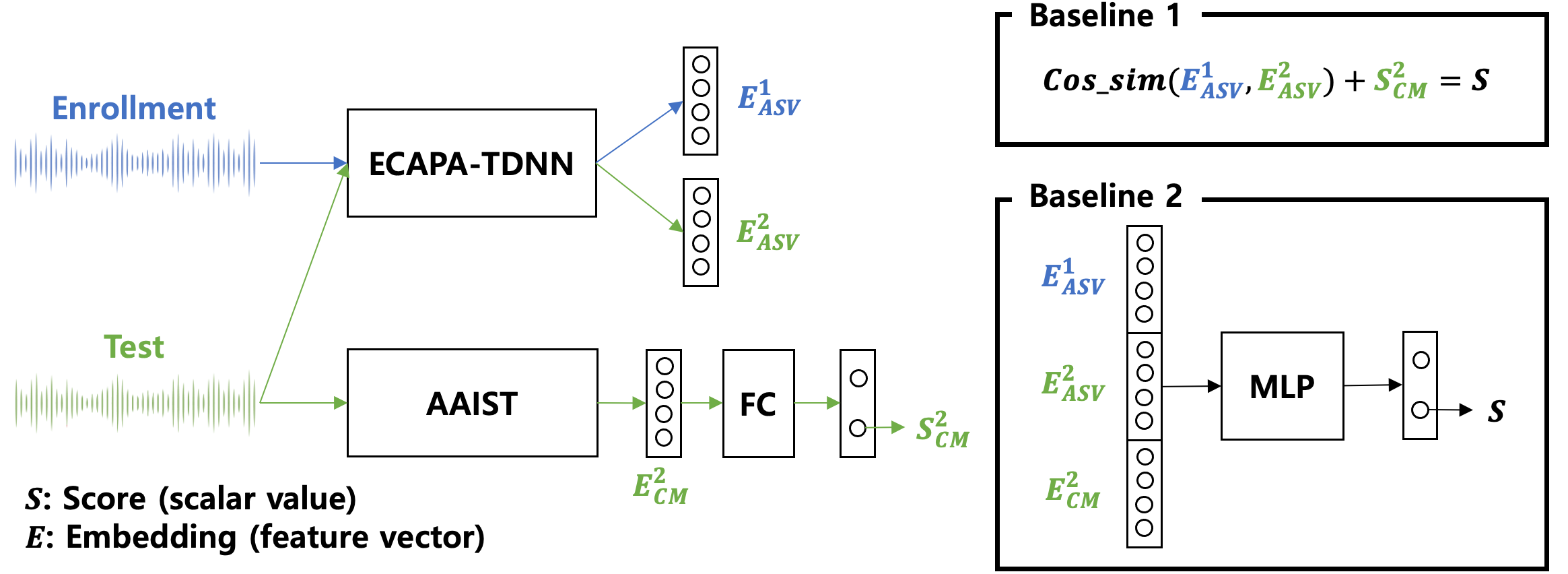}
\vspace{-0.2em}
\caption{
    The structure of the baselines proposed by the SASV challenge organizers \cite{jung2022sasv}. 
    Baseline1, score-sum ensemble system, derives SASV score ($\mathbf{S}$) by summing ASV and cm scores. 
    On the other hand, baseline2, back-end model ensemble system, fuses three embedding by multi-layer perceptron and output ($\mathbf{S}$). 
    The ECAPA-TDNN and AAIST are pre-trained subsystems which performs ASV and CM. 
}
\label{figure:baseline}
\end{center}
\vspace{-2em}
\end{figure*}

\section{Spoofing-Aware Speaker Verification (SASV) Challenge 2022}
\label{section:sasv2022}
The Spoofing-Aware Speaker Verification (SASV) challenge 2022 initiated spearheads research in automatic speaker verification (ASV) which contemplates spoofing countermeasure (CM). 
Challenge aims to improve the robustness of the ASV system to spoofing attacks by optimizing CM and ASV systems operating in tandem and, finally, boosting the development of single integrated systems. 
To facilitate research, challenge organizers provided datasets, baselines, and metrics, and we describe them in this section.

\begin{table}[!hb]
\centering
\caption{
    Description of the ASVspoof 2019 LA dataset \cite{todisco2019asvspoof}.
}
\resizebox{0.9\linewidth}{!}{%
    \begin{tabular}{c|cc|cc}
        \Xhline{2\arrayrulewidth}
        \multirow{2}{*}{Partition} & \multicolumn{2}{c}{Speaker} & \multicolumn{2}{c}{Utterance} \\ \cline{2-5} 
         & Male & Female & Bonafide & Spoof \\ 
        \hline
        Train & 8 & 12 & 2,580 & 22,800 \\
        Development & 4 & 6 & 2,548 & 22,296 \\
        Evaluation & 21 & 27 & 7,355 & 63,882 \\ 
        \Xhline{2\arrayrulewidth}
    \end{tabular}
    \label{table:dataset}
    \vspace{-2cm}
}
\end{table}

\begin{table}[!ht]
\centering
\caption{
    Description of SV, SPF, and SASV EER. 
    “+” and “-”  denotes \textit{target} and \textit{non-target} class, and a blank denotes not used class. 
    SASV-EER is calculated using the full trials, while others is calculated using subset of trials. 
}
 
\resizebox{0.9\linewidth}{!}{%
    \begin{tabular}{c | c c c}
    \Xhline{2\arrayrulewidth}
        & Target speaker & Non-target speaker & Spoof\\
    \hline
        SV EER   & +      & -          &       \\
        SPF EER  & +      &            & -     \\
        SASV EER & +      & -          & -     \\
    \Xhline{2\arrayrulewidth}
    \end{tabular}
    \label{table:metric}
    \vspace{-2cm}
}
\end{table}

\subsection{Dataset}
This year, the challenge focused on logical access (LA) scenarios that contain speech synthesis and voice conversion attacks. 
Therefore, the ASV spoof 2019 LA partitions \cite{wang2020asvspoof, todisco2019asvspoof}, a dataset designed for the LA task of ASVspoof 2019, is mainly used in the challenge. 
It consists of bonafide speech collected from VCTK corpus \cite{yamagishi2019cstr} and spoofed speech generated using 17 different text-to-speech (TTS) and voice conversion (VC) system. 
There are three partitions in ASV spoof 2019 LA dataset; train set, development set, and evaluation set; no overlap between partitions, and evaluation partitions are created with different TTS, VC systems than train and development partitions. 
Details are shown in Table \ref{table:dataset}. 
SASV challenge organizers also provide development and evaluation protocols listing the \textit{target} and \textit{non-target} trials.

\subsection{Baseline}
Challenge Organizers presented two baselines; a score-sum ensemble system (baseline1) and a back-end model ensemble system (baseline2). 
Both systems involve two pre-trained subsystems, ECAPA-TDNN \cite{desplanques2020ecapa} and AAIST \cite{jung2021aasist}, a conventional ASV and CM system. 
ECAPA-TDNN is based on Res2Net with squeeze-excitation modules and outputs speaker embedding from input spectrogram. 
The ASV score is obtained by calculating the cosine similarity between test utterance and enrollment utterance embedding. 
Meanwhile, AAIST is based on an integrated spectral-temporal graph attention network and uses raw-waveform as an input. 
Although this system was designed to output CM scores directly, the challenge organizer also utilized feature vectors fed to the output layer of AAIST.

\begin{figure}[ht!]
\begin{center}
    \centering
    \includegraphics[width=0.9\linewidth]{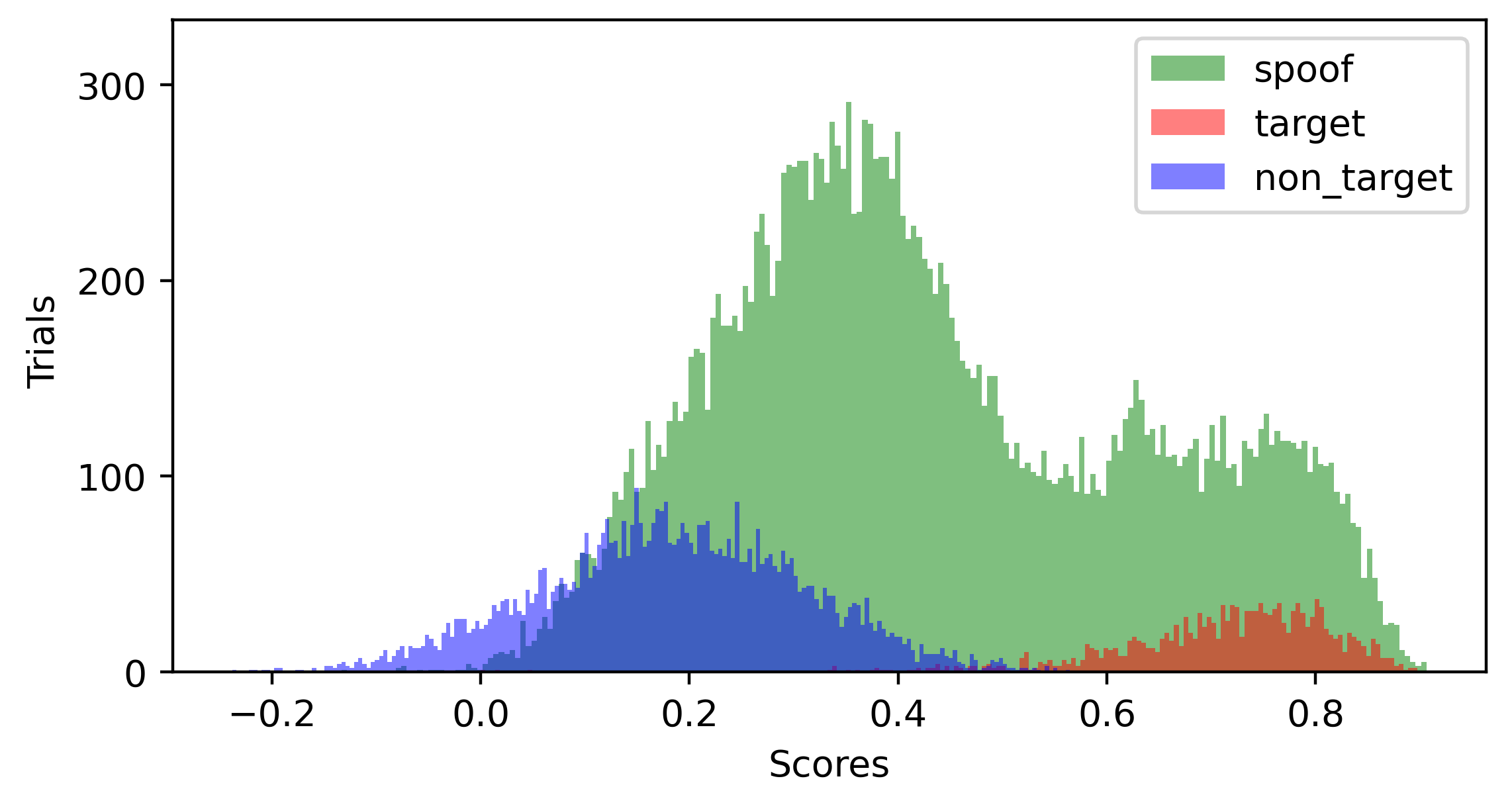}
\caption{
    Distribution of the ECAPA-TDNN ASV scores for all development trials. 
    We used all trials, including spoofing, to check how the ASV system behaves upon input of spoofed utterances. 
    The red, blue, and green denotes distribution of \textit{target}, \textit{non-target}, \textit{and spoofed}. 
}
\label{figure:distribution}
\end{center}
\vspace{-3em}
\end{figure}

Figure \ref{figure:baseline} describes the structure of baselines. 
As shown in Figure \ref{figure:baseline}, baseline1 simply sums the ASV and CM score. 
Therefore, it does not requires any training nor fine-tuning. 
On the other hand, baseline2 is a multi-layer perceptron with three 1024 hidden units that fuses three embeddings: two extracted from test and enrollment utterances using the ECAPA-TDNN system; the last one extracted from the same test utterance using the AAIST system. 
Unlike baseline1, baseline2 requires training using the ASV spoof 2019 LA train partition.

\begin{figure}[!ht]
\begin{center}
    \centering
    \includegraphics[width=0.9\linewidth]{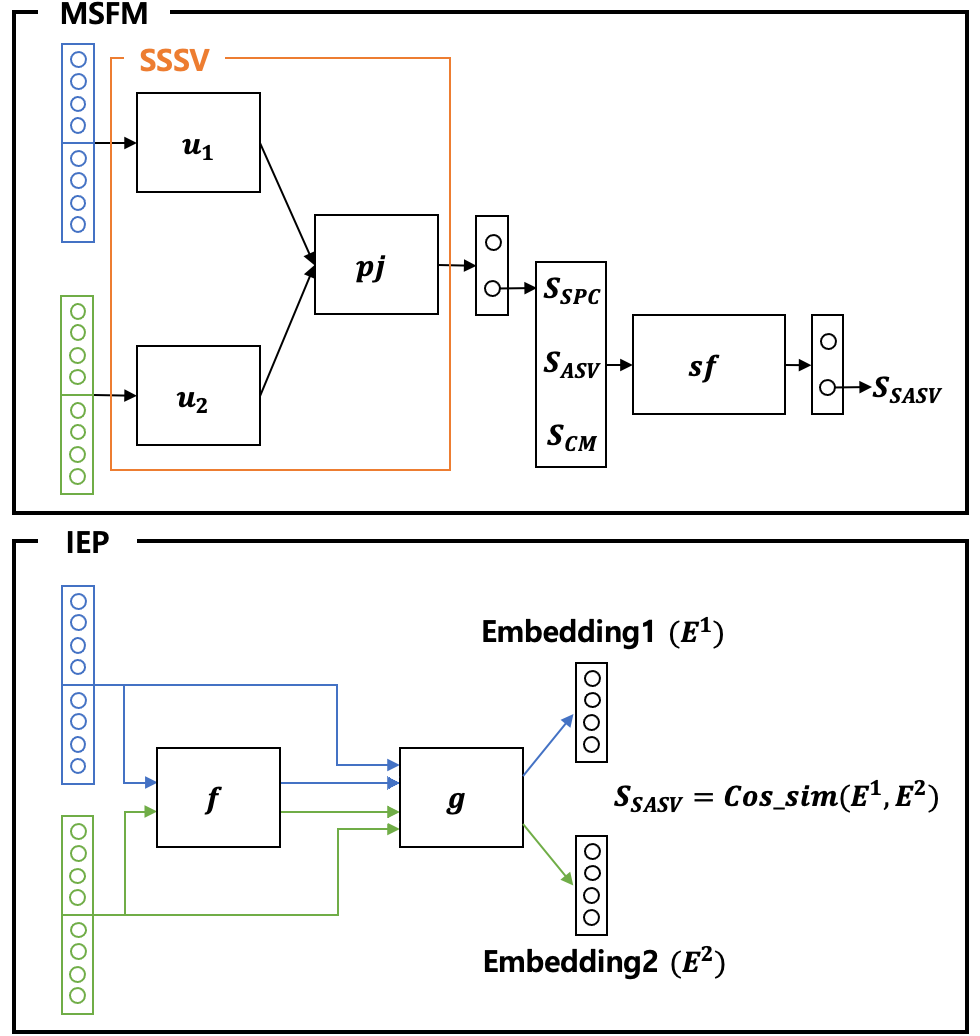}
\caption{
    Description of the structure of the proposed frameworks. Blue and green boxes are the feature vectors that concatenated ASV and CM embeddings extracted from enrollment and test utterances. 
    The $\mathbf{S_{task}}$ means the score of each $\mathbf{task}$. 
}
\label{figure:proposed}
\end{center}
\vspace{-2em}
\end{figure}

\subsection{Evaluation metrics} 
The primary metric of the challenge is SASV equal error rate (EER), but the SASV-EER does not distinguish between different speaker attempts and spoofing attempts. 
Therefore challenge organizers also embraced speaker verification EER (SV-EER) and anti-spoofing EER (SPF-EER). 
Table \ref{table:metric} shows the description of EER. 
The SASV-EER is produced using the full set of challenge public trials. 
In contrast, as shown in Table \ref{table:metric}, challenge organizers present SV-EER and CM-EER excluding some trials because some trials are affected by both tasks. 
For example, suppose the SASV system rejects non-target speaker's spoofed utterance. 
In that case, it is hard to know if this is due to speaker-to-speaker discrepancies, spoofed utterances, or both.

\section{Proposed method}
\label{section:proposed_method}
Inspired by the baselines in SASV challenge, we adopt back-end systems to integrate ASV and CM systems. 
The proposed framework configuration is shown in Figure \ref{figure:proposed} and the structure of each module is described in Table \ref{table:structures}. 
As depicted in Figure \ref{figure:proposed}, we proposed 2 frameworks, multi-layer perceptron score fusion model (MSFM) and integrated embedding projector (IEP), to compare score-base fusion strategy and embedding-base fusion strategy. 
In this section, we introduce our proposed frameworks in order to MSFM and IEP.

\subsection{Multi-layer perceptron score fusion model}
The proposed multi-layer perceptron score fusion model (MSFM) is based on the motivation from several observations of the experiments. 
Figure \ref{figure:distribution} shows the distribution of the ECAPA-TDNN’s ASV score for all development trials. 
The distribution of target (green) and non-target (blue) are clearly separated, which means the ASV subsystem reliably performs in bonafide scenarios. 
However, the distribution of spoofed utterances (green) is widely dispersed, overlaying both the target and non-target distributions. 
This phenomenon is due to the degradation of the discrimination of the ASV system in the spoofing scenarios because conventional ASV systems have never seen spoofing scenarios in the learning process. 
When integrating the results of the two systems, this performance degradation can have a negative impact because only the ASV and CM scores affect the final SASV score.

\begin{table}[!ht]
\centering
\caption{
    Description of the structure of the modules used in the proposed framework. 
    The left column denotes the modules of SSSV and MSFM, and the right column shows the modules of IEP. 
    SSSV consists of the two embedding fusion blocks ($u1$, $u2$) and a score calculation block ($pj$). 
    ELU is the exponential linear unit activation function proposed in \cite{clevert2015fast}.
}

\resizebox{0.9\linewidth}{!}{%
    \label{table:structures}
    \begin{tabular}{c c | c c}
    \Xhline{2\arrayrulewidth}
    \textbf{Layer} & \textbf{Structure} & \textbf{Layer} & \textbf{Structure}\\
    \Xhline{2\arrayrulewidth}
        $u_1$ & 
        \begin{tabular}{c}
            FC($352 \times 128$)\\
            ELU\\
            FC($128 \times 128$)\\
            ELU\\
            FC($128 \times 64$)\\
            ELU\\
            FC($64 \times 160$)\\
            \end{tabular} & 
        $f$ & 
        \begin{tabular}{c}
            FC($352 \times 256$)\\
            ELU\\
            FC($256 \times 256$)\\
            ELU\\
            FC($256 \times 128$)\\
            ELU\\
            \end{tabular} \\
    \midrule
        $u_2$ & 
        \begin{tabular}{c}
            FC($352 \times 128$)\\
            ELU\\
            FC($128 \times 128$)\\
            ELU\\
            FC($128 \times 64$)\\
            ELU\\
            FC($64 \times 160$)\\
            \end{tabular} & 
        $g$ &  FC($480 \times 128$) \\
    \midrule
        $pj$ & 
        \begin{tabular}{c}
            FC($320 \times 128$)\\
            ELU\\
            FC($128 \times 64$)\\
            ELU\\
            FC($64 \times 2$)\\
            \end{tabular} & 
        \multicolumn{2}{c}{-} \\
    \midrule
        $sf$ & 
        \begin{tabular}{c}
            FC($2 or 3 \times 16$)\\
            ELU\\
            FC($16 \times 16$)\\
            ELU\\
            FC($16 \times 2$)\\
            \end{tabular} & 
        \multicolumn{2}{c}{-} \\
    \Xhline{2\arrayrulewidth}
    \end{tabular}
}
\vspace{-0.5cm}
\end{table}

\begin{table*}[!ht]
\centering
\caption{
    Experimental results (EER, \%) of speaker verification (SV), anti-spoofing (SPF), and spoofing-aware speaker verification (SASV) tasks for the SASV 2022 challenge development and evaluation protocols. 
}
\label{table:performance}
\resizebox{0.7\linewidth}{!}{
    \begin{tabular}{lcccccc}
    \Xhline{2\arrayrulewidth}
     & \multicolumn{2}{c}{SV-EER} & \multicolumn{2}{c}{SPF-EER} & \multicolumn{2}{c}{SASV-EER} \\ \cline{2-7}
     & Dev & Eval & Dev & Eval & Dev & Eval \\ 
     \hline
    ECAPA-TDNN & 1.88 & 1.63 & 20.30 & 30.75 & 17.38 & 23.83 \\
    Baseline1 \cite{jung2022sasv} & 32.88 & 35.32 & \textbf{0.06} & 0.67 & 13.07 & 19.31 \\
    Baseline2 \cite{jung2022sasv} & 12.87 & 11.48 & 0.13 & 0.78  & 4.85 & 6.37 \\ \hline
    MSFM & \textbf{1.11} & \textbf{0.73} & 0.13 & \textbf{0.43} & 0.67 & \textbf{0.56} \\
    MSFM (without SSSV score) & 1.3 & 0.96 & 0.17 & 0.44 & \textbf{0.60} & 0.72 \\
    IEP & 2.51 & 1.58 & 2.7 & 1.12 & 1.55 & 1.32 \\
    \Xhline{2\arrayrulewidth}
    \end{tabular}
}
\vspace{-0.4cm}
\end{table*}

Therefore, to compensate the ASV score, we further utilized a spoofing scenario speaker verification (SSSV) system that learns ASV in spoofing scenarios. 
The SSSV model is fed by the embeddings of the ASV and CM subsystems and output a speaker verification score. 
It consists of two embedding fusion blocks ($u1$, $u2$) and a score calculation block ($pj$). 
ASV and CM embeddings extracted from test and enrollment utterances are concatenated and fed to $u1$. 
In the same way, the embeddings extracted from test utterances are fed to $u2$. 
$u1$ and $u2$ combine the speaker and spoofing information in the embeddings and output 160-dimensional feature vectors. 
Then the block $pj$ converts the feature vector into a speaker verification score vector. 
The first node of the output score layer represents the non-target score, and the second node represents the target score. 
We used only the value of the second node as a SSSV score. 
For the score fusion, we used the multi-layer perceptron ($sf$) which has been widely used for classification. 
The structure of the MSFM is described in the left column of Table \ref{table:structures}. 
It consists of three fully-connected (FC) layers and two exponential linear unit (ELU) layers. 

SSSV and $sf$ are trained simultaneously using categorical cross entropy (CCE) criterion. The loss is calculated as follows. 
Here, $t$, $l$ are the ground truths of ASV and SASV, and $s$, $v$ denote score vectors output from SSSV and $sf$, which has 2 nodes.
\begin{gather}
    \mathcal{L}_{SSSV} = - \sum^2_i t_i\log(softmax(s)_i), \\
    \mathcal{L}_{sf} = - \sum^2_i l_i\log(softmax(v)_i), \\
    \mathcal{L}_{TOTAL} = \mathcal{L}_{SSSV} + \mathcal{L}_{sf}.
\end{gather}

\begin{table}[!ht]
\centering
\caption{
    Description of the training pairs. 
    The training pair consists of two utterances and one target or non-target label. 
    One utterance is always bonafide, and the other utterance is bonafide or spoofed. 
    “+” and “-”  denote \textit{target} and \textit{non-target} classes. 
}
\label{table:train_pair}
\begin{tabular}{c|cc}
    \Xhline{2\arrayrulewidth}
    & Bonafide & Spoofed \\ 
    \hline
    Same speaker     & +        & -       \\ 
    Different speaker & -        & -       \\ 
    \Xhline{2\arrayrulewidth}
\end{tabular}
\vspace{-0.3cm}
\end{table}

Training the MSFM system requires the training pairs which contain an enrollment utterance, a test utterance, and a SASV label. 
However, ASV spoof 2019 LA training set provides only the utterances, speaker identities, and spoof keys. 
Therefore, it was necessary to construct training pairs. 
We designed a training pairs based on four scenarios as shown in Table \ref{table:train_pair}. 
The ratios for each scenario are (bonafide same speaker):(bonafice different speaker):(spoofed same speaker):(spoofed different speaker) = 3:1.66:1:1, and we used total 2000 samples per epoch.

\subsection{Integrated embedding projector}
We thought embeddings had higher potential than scores because embeddings contained more information than scores.
Therefore, we designed the back-end model called Integrated Embedding Projector (IEP) that transforms ASV and CM embeddings into SASV embeddings using a metric learning. 
As shown in Figure \ref{figure:proposed}, IEP consists of two modules, $f$ and $g$, and the structure of each module is described in Figure \ref{figure:proposed}. 
The feedforward process of IEP is as follows: 
\begin{equation}
    z = g(concat(f(x, y), x, y)),
\end{equation}
where $z$ denotes the output of the IEP (SASV embedding) and
$x$, $y$ denote the ASV, CM embeddings. 
We iteratively fed $x$ and $y$ embedding to the $g$ model to re-aggregate information of each task that can be distorted during the embedding fusion process. 
The IEP model is trained using cosine similarity-based triplet loss \cite{weinberger2009distance} to explore the embedding space for SASV. 
Considering the characteristics of the SASV task, we construct a triplet for the training as follows:
\begin{itemize}
    \item Anchor ($A_{i}$): $i$th speaker's bonafide embeddings.
    \item Positive pair ($P_{i}$): $i$th speaker's bonafide embeddings which are extracted from different utterances other than the anchor. 
    \item Negative pair ($N$): $i$th speaker's any spoof embeddings or other speaker's any bonafide embeddings. 
\end{itemize}

Therefore, our proposed IEP is trained to optimize the triplet loss as follows: 
\begin{equation}
    \mathcal{L}_{TRIPLET} = \frac{1}{c}\sum^c_{i=1} max(0, cos(A_{i}, N) - cos(A_{i}, P_{i}) + m),
\end{equation}
where, $c$ is the number of triplets per single mini-batch and $m$ is the margin, set to 0.5. 
Finally, SASV score is calculated based on the cosine similarity.

\begin{figure}[ht!]
\begin{center}
    \centering
    \includegraphics[width=0.9\linewidth]{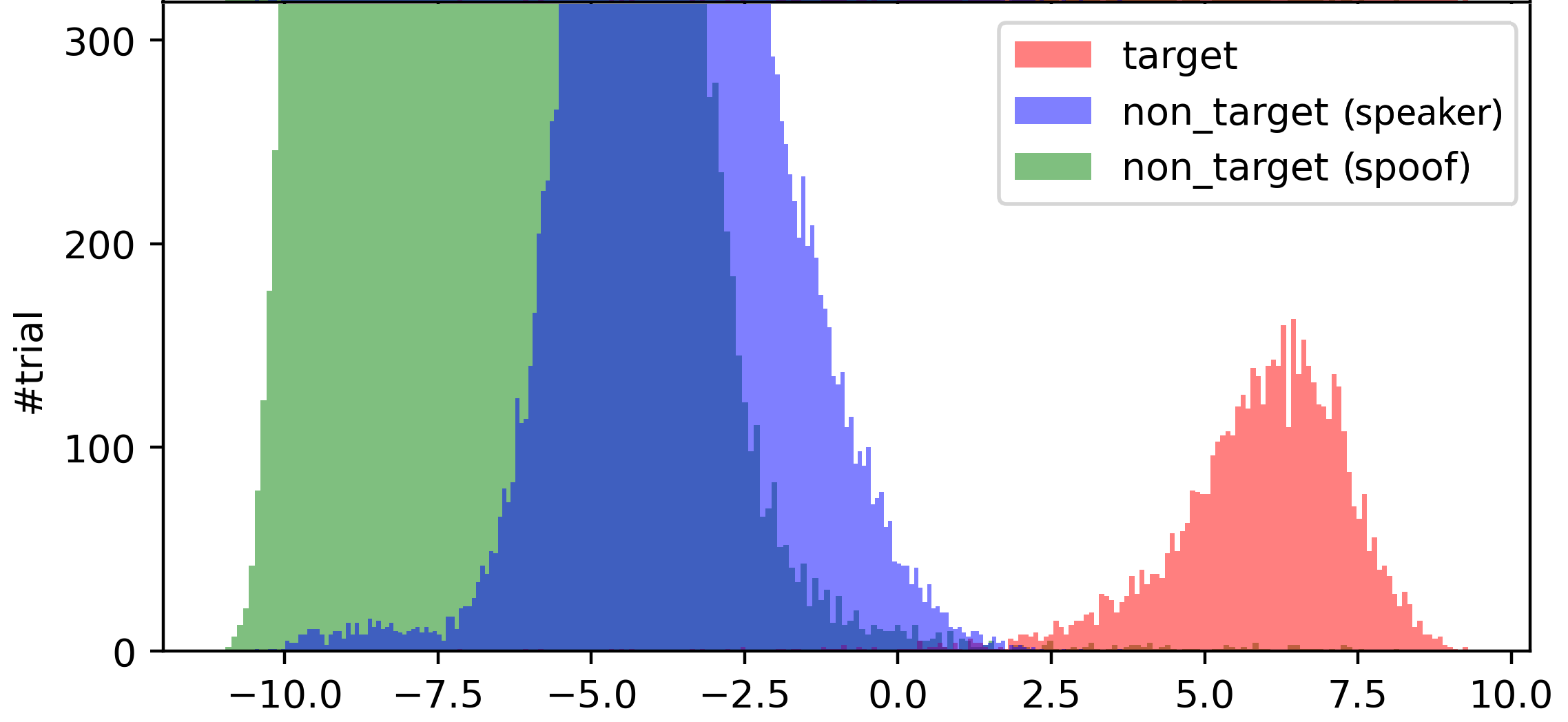}
\caption{
    Distribution of the proposed MSFM scores for all evaluation trials. 
    The red depicts distribution of \textit{target}, and blue and green indicate \textit{non-target}.
}
\label{figure:dist2}
\end{center}
\vspace{-3em}
\end{figure}

\section{Results}
\label{section:results}
SASV research is in its infancy, and the SASV Challenge was held for the first time in this year. 
Therefore, we compared our systems with the challenge baselines, as depicted in Table \ref{table:performance}. 

Baseline1, which adds the ASV and CM scores, reported the SASV-EER of 19\%. 
This is poor performance considering the performance of subsystems (ECAPA-TDNN and AAIST each showed less than 2\% EER in their fields). 
The proposed MSFM, meanwhile, achieved EER of 0.56\% for the SASV evaluation protocol. 
This is further improved performance than baseline1. 
Also, when comparing MSFM and MSFM (without SSSV score) results, using the SSSV score is more effective. 
Through this results, we can confirm that SSSV score might help to improve ASV and SASV performance. 
Figure \ref{figure:dist2} shows the distribution of the proposed MSFM scores for all SASV evaluation trials, and we can notice that the \textit{target} and the \textit{non-target} are well separated. 

On the other hand, the proposed IEP showed an EER of 1.32\%. 
That is a poor performance than the MSFM, but it was improved compared to baseline2, same embedding fusion method as IEP. 
It is analyzed that the proposed IEP model was well trained in the embedding space suitable for SASV through metric learning.

\section{Conclusion}
\label{section:conclusion}
In this work, we propose MSFM and IEP systems to integrate ASV and CM systems. 
MSFM converges ASV, CM, and the presented SSSV score with multi-layer perceptron. 
On the other hand, IEP casts ASV and CM embeddings into SASV embedding and produces SASV score by cosine similarity. 
MSFM and IEP systems are demonstrated the EER of 0.56\% and 1.32\%, respectively. 
SASV research is in its infancy, and in this study, we conducted research in a direction that does not significantly modify the conventional ASV and CM systems. 
However, by optimizing two systems together, the system might offer superior performance. 
Therefore, in future studies, we plan to build a fully integrated SASV system. 

\section{Acknowledgement}
This research was supported by Basic Science Research Program through the National Research Foundation of Korea (NRF) funded by the Ministry of Science, ICT Future Planning (2020R1A2C1007081).

\bibliographystyle{IEEEtran}

\bibliography{references}

\begin{thebibliography}{10}
\providecommand{\url}[1]{#1}
\csname url@samestyle\endcsname
\providecommand{\newblock}{\relax}
\providecommand{\bibinfo}[2]{#2}
\providecommand{\BIBentrySTDinterwordspacing}{\spaceskip=0pt\relax}
\providecommand{\BIBentryALTinterwordstretchfactor}{4}
\providecommand{\BIBentryALTinterwordspacing}{\spaceskip=\fontdimen2\font plus
\BIBentryALTinterwordstretchfactor\fontdimen3\font minus
  \fontdimen4\font\relax}
\providecommand{\BIBforeignlanguage}[2]{{%
\expandafter\ifx\csname l@#1\endcsname\relax
\typeout{** WARNING: IEEEtran.bst: No hyphenation pattern has been}%
\typeout{** loaded for the language `#1'. Using the pattern for}%
\typeout{** the default language instead.}%
\else
\language=\csname l@#1\endcsname
\fi
#2}}
\providecommand{\BIBdecl}{\relax}
\BIBdecl

\bibitem{variani2014deep}
E.~Variani, X.~Lei, E.~McDermott, I.~L. Moreno, and J.~Gonzalez-Dominguez,
  ``Deep neural networks for small footprint text-dependent speaker
  verification,'' in \emph{2014 IEEE international conference on acoustics,
  speech and signal processing (ICASSP)}.\hskip 1em plus 0.5em minus
  0.4em\relax IEEE, 2014, pp. 4052--4056.

\bibitem{dehak2010front}
N.~Dehak, P.~J. Kenny, R.~Dehak, P.~Dumouchel, and P.~Ouellet, ``Front-end
  factor analysis for speaker verification,'' \emph{IEEE Transactions on Audio,
  Speech, and Language Processing}, vol.~19, no.~4, pp. 788--798, 2010.

\bibitem{snyder2018x}
D.~Snyder, D.~Garcia-Romero, G.~Sell, D.~Povey, and S.~Khudanpur, ``X-vectors:
  Robust dnn embeddings for speaker recognition,'' in \emph{2018 IEEE
  international conference on acoustics, speech and signal processing
  (ICASSP)}.\hskip 1em plus 0.5em minus 0.4em\relax IEEE, 2018, pp. 5329--5333.

\bibitem{thienpondt2021idlab}
J.~Thienpondt, B.~Desplanques, and K.~Demuynck, ``The idlab voxsrc-20
  submission: Large margin fine-tuning and quality-aware score calibration in
  dnn based speaker verification,'' in \emph{ICASSP 2021-2021 IEEE
  International Conference on Acoustics, Speech and Signal Processing
  (ICASSP)}.\hskip 1em plus 0.5em minus 0.4em\relax IEEE, 2021, pp. 5814--5818.

\bibitem{heo2020clova}
H.~S. Heo, B.-J. Lee, J.~Huh, and J.~S. Chung, ``Clova baseline system for the
  {VoxCeleb} speaker recognition challenge 2020,'' \emph{arXiv preprint
  arXiv:2009.14153}, 2020.

\bibitem{shim2021graph}
H.-j. Shim, J.~Heo, J.-h. Park, G.-h. Lee, and H.-J. Yu, ``Graph attentive
  feature aggregation for text-independent speaker verification,'' \emph{arXiv
  preprint arXiv:2112.12343}, 2021.

\bibitem{wang2020asvspoof}
X.~Wang, J.~Yamagishi, M.~Todisco, H.~Delgado, A.~Nautsch, N.~Evans,
  M.~Sahidullah, V.~Vestman, T.~Kinnunen, K.~A. Lee \emph{et~al.}, ``Asvspoof
  2019: A large-scale public database of synthesized, converted and replayed
  speech,'' \emph{Computer Speech \& Language}, vol.~64, p. 101114, 2020.

\bibitem{wu2015asvspoof}
Z.~Wu, T.~Kinnunen, N.~Evans, J.~Yamagishi, C.~Hanil{\c{c}}i, M.~Sahidullah,
  and A.~Sizov, ``Asvspoof 2015: the first automatic speaker verification
  spoofing and countermeasures challenge,'' in \emph{Sixteenth annual
  conference of the international speech communication association}, 2015.

\bibitem{wu2017asvspoof}
Z.~Wu, J.~Yamagishi, T.~Kinnunen, C.~Hanil{\c{c}}i, M.~Sahidullah, A.~Sizov,
  N.~Evans, M.~Todisco, and H.~Delgado, ``Asvspoof: the automatic speaker
  verification spoofing and countermeasures challenge,'' \emph{IEEE Journal of
  Selected Topics in Signal Processing}, vol.~11, no.~4, pp. 588--604, 2017.

\bibitem{delgado2021asvspoof}
H.~Delgado, N.~Evans, T.~Kinnunen, K.~A. Lee, X.~Liu, A.~Nautsch, J.~Patino,
  M.~Sahidullah, M.~Todisco, X.~Wang \emph{et~al.}, ``Asvspoof 2021: Automatic
  speaker verification spoofing and countermeasures challenge evaluation
  plan,'' \emph{arXiv preprint arXiv:2109.00535}, 2021.

\bibitem{kinnunen2020tandem}
T.~Kinnunen, H.~Delgado, N.~Evans, K.~A. Lee, V.~Vestman, A.~Nautsch,
  M.~Todisco, X.~Wang, M.~Sahidullah, J.~Yamagishi \emph{et~al.}, ``Tandem
  assessment of spoofing countermeasures and automatic speaker verification:
  Fundamentals,'' \emph{IEEE/ACM Transactions on Audio, Speech, and Language
  Processing}, vol.~28, pp. 2195--2210, 2020.

\bibitem{jung2022sasv}
J.-w. Jung, H.~Tak, H.-j. Shim, H.-S. Heo, B.-J. Lee, S.-W. Chung, H.-G. Kang,
  H.-J. Yu, N.~Evans, and T.~Kinnunen, ``Sasv challenge 2022: A spoofing aware
  speaker verification challenge evaluation plan,'' \emph{arXiv preprint
  arXiv:2201.10283}, 2022.

\bibitem{eskimez2018front}
S.~E. Eskimez, P.~Soufleris, Z.~Duan, and W.~Heinzelman, ``Front-end speech
  enhancement for commercial speaker verification systems,'' \emph{Speech
  Communication}, vol.~99, pp. 101--113, 2018.

\bibitem{kolboek2016speech}
M.~Kolboek, Z.-H. Tan, and J.~Jensen, ``Speech enhancement using long
  short-term memory based recurrent neural networks for noise robust speaker
  verification,'' in \emph{2016 IEEE spoken language technology workshop
  (SLT)}.\hskip 1em plus 0.5em minus 0.4em\relax IEEE, 2016, pp. 305--311.

\bibitem{heymann2017beamnet}
J.~Heymann, L.~Drude, C.~Boeddeker, P.~Hanebrink, and R.~Haeb-Umbach,
  ``Beamnet: End-to-end training of a beamformer-supported multi-channel asr
  system,'' in \emph{2017 IEEE International Conference on Acoustics, Speech
  and Signal Processing (ICASSP)}.\hskip 1em plus 0.5em minus 0.4em\relax IEEE,
  2017, pp. 5325--5329.

\bibitem{bai2021universal}
Z.~Bai and B.~Hu, ``A universal bert-based front-end model for mandarin
  text-to-speech synthesis,'' in \emph{ICASSP 2021-2021 IEEE International
  Conference on Acoustics, Speech and Signal Processing (ICASSP)}.\hskip 1em
  plus 0.5em minus 0.4em\relax IEEE, 2021, pp. 6074--6078.

\bibitem{todisco2019asvspoof}
M.~Todisco, X.~Wang, V.~Vestman, M.~Sahidullah, H.~Delgado, A.~Nautsch,
  J.~Yamagishi, N.~Evans, T.~Kinnunen, and K.~A. Lee, ``Asvspoof 2019: Future
  horizons in spoofed and fake audio detection,'' \emph{arXiv preprint
  arXiv:1904.05441}, 2019.

\bibitem{yamagishi2019cstr}
J.~Yamagishi, C.~Veaux, K.~MacDonald \emph{et~al.}, ``Cstr vctk corpus: English
  multi-speaker corpus for cstr voice cloning toolkit (version 0.92),'' 2019.

\bibitem{desplanques2020ecapa}
B.~Desplanques, J.~Thienpondt, and K.~Demuynck, ``Ecapa-tdnn: Emphasized
  channel attention, propagation and aggregation in tdnn based speaker
  verification,'' in \emph{Interspeech2020}.\hskip 1em plus 0.5em minus
  0.4em\relax International Speech Communication Association (ISCA), 2020, pp.
  3830--3834.

\bibitem{jung2021aasist}
J.-w. Jung, H.-S. Heo, H.~Tak, H.-j. Shim, J.~S. Chung, B.-J. Lee, H.-J. Yu,
  and N.~Evans, ``Aasist: Audio anti-spoofing using integrated spectro-temporal
  graph attention networks,'' \emph{arXiv preprint arXiv:2110.01200}, 2021.

\bibitem{clevert2015fast}
D.-A. Clevert, T.~Unterthiner, and S.~Hochreiter, ``Fast and accurate deep
  network learning by exponential linear units (elus),'' \emph{arXiv preprint
  arXiv:1511.07289}, 2015.

\bibitem{weinberger2009distance}
K.~Q. Weinberger and L.~K. Saul, ``Distance metric learning for large margin
  nearest neighbor classification.'' \emph{Journal of machine learning
  research}, vol.~10, no.~2, 2009.

\end{thebibliography}

\end{document}